\documentstyle[twoside,12pt,array,fleqn,epsf]{ioplppt}
\catcode `\@=11
\def\numberbysection{\@addtoreset{equation}{section}
         \renewcommand{\theequation}{\thesection.\arabic{equation}}}

\def\be{\begin{equation}}
\def\ee{\end{equation}}
\def\bd{\begin{displaymath}}
\def\ed{\end{displaymath}}
\newcommand{\ba}{\begin{eqnarray}}
\newcommand{\ea}{\end {eqnarray}}
\newcommand{\nn}{\nonumber}

\def\BCT{\,\hbox{\hbox to -3pt{\vrule height 6.5pt width .2pt\hss}\rm
C}}
\def\BRT{\,\hbox{\hbox to -1pt{\vrule height 7.4pt width .2pt\hss}\rm
R}}
\def \eins{{\rm 1 \kern-2.8pt I }}

\def\G{\Gamma}

\def\m{\mu}
\def\n{\nu}
\def\rh{\rho}

\def\s{\sigma}

\def\Q{{\cal Q}}

\def\half{\frac{1}{2}}

\numberbysection
\begin{document}
\pagestyle{plain}
\vspace* {10mm}
\begin{center}
\large
{\bf Spectra of non-hermitian quantum spin chains describing
boundary induced phase transitions} 
\end{center}
\begin{center}
\normalsize
       Ulrich Bilstein\footnote{bilstein@theoa1.physik.uni-bonn.de} and
       Birgit Wehefritz\footnote{birgit@theoa1.physik.uni-bonn.de}
        \\[1cm]
    {\it Universit\"{a}t Bonn,
                    Physikalisches Institut \\ Nu\ss allee 12,
                    D-53115 Bonn, Germany}\\[14mm]
{\bf Abstract}
\end{center}
\small
\noindent
The spectrum of the non-hermitian asymmetric XXZ-chain with additional
non-diagonal boundary terms is studied. The lowest lying
eigenvalues are determined numerically. For the ferromagnetic and completely
asymmetric chain that corresponds to a reaction-diffusion model
with input and outflow of particles
the smallest energy gap correponding directly to the
inverse of the temporal correlation length shows the same properties as
the spatial correlation length of the stationary state. 
For the antiferromagnetic
chain with both boundary terms, we find a conformal invariant spectrum
where the partition function corresponds to the one of a Coulomb gas
with only magnetic charges  shifted by a purely imaginary 
and a lattice-length dependent constant. Similar results are obtained
by studying a toy model that can be diagonalized analytically in terms of
free fermions.
\rule{5cm}{0.2mm}
\begin{flushleft}
\parbox[t]{3.5cm}{\bf PACS numbers:}
\parbox[t]{12.5cm}{05.70.Ln, 64.60.Ht, 64.90.+b, 75.10.Jm}
\end{flushleft}
\normalsize

\newpage
\pagestyle{plain}

\section{Introduction}
In the study of reaction-diffusion processes
non-hermitian chains
appear
in a natural way. However, 
their properties  have not yet
been studied extensively.
For instance, the effect of boundary conditions on the spectrum of non-hermitian
Hamiltonians is unknown, although a rich and surprising structure arises here.
In this article we study an asymmetric XXZ-chain with modified
boundary terms and we show that these boundary terms give rise to new 
and interesting features.

In this article, two examples are treated that both show
the appearance of boundary-induced phase transitions. As a first example,
we study the Hamiltonian of the asymmetric diffusion model on a one-dimensional
lattice with open boundaries, injection 
and ejection of particles at the edges of the chain.
For this model, all known results are valid only for the stationary state.
The time evolution is studied for the first time in this article. 
In this model, particles of one species can diffuse on the lattice with
length $N$ with a rate $p$ to the right and $q$ to the left. 
Particles are injected at the left boundary with a rate $\alpha$
 and extracted with a rate
$\beta$ at the right boundary. The Hamiltonian of this model
is non-hermitian and is given
by the following expression:
\be
H=-\sum_{j=1}^{N-1}\left[q\sigma_j^-\sigma_{j+1}^+
 +p\sigma_j^+\sigma_{j+1}^-+ \frac{p+q}{4}(\sigma^z_j\sigma^z_{j+1}
-1) + \frac{q-p}{4}(\sigma_{j+1}^z-\sigma_j^z)\right]+B_1+B_N
\label{Ham}
\ee
with 
\ba
B_1 & =& \quad\frac{\alpha}{2}(\sigma_1^z-2\sigma_1^-
+1)\nn\\
B_N & = & -\frac{\beta}{2}(\sigma_N^z +2\sigma_N^+
-1)
\label{randein}
\ea
The bulk term of this Hamiltonian corresponds to the asymmetric
XXZ-Hamiltonian with anisotropy $\Delta = \frac{p+q}{2}$.
This model and versions of it have been studied extensively.
The total asymmetric diffusion
model (with $q=0$) was first defined by Derrida, Domany and Mukamel \cite{muk}
who found
the phase diagram for the current in the stationary state as a function of 
$\alpha$ and $\beta$ by application of mean-field theory. The ground state of
the model can
be determined exactly by a matrix product formulation \cite{Schutz_Domany,DEHP}.
Two massive and one 
massless phase have been found for different regions of the $\alpha$-$\beta$ 
plane. The massive phases are separated by a line on which the spectrum
is massless. In this context massless means that the spatial correlation
length diverges, which implies an algebraic falloff of the 
spatial concentration
profile. On the other hand, massive means that the spatial correlation length
remains finite. 
The phase diagram for the partially asymmetric system ($p \neq q \neq 0$)
was presented for the first time in \cite{Sandow}. Essler and Rittenberg
\cite{Fab} calculated correlation functions for this system with the
help of the matrix product formulation. 

The case of periodic boundary conditions
 has previously been studied by Gwa and Spohn \cite{Gwa_Spohn} for $q=0$
and $p=1$ using Bethe-Ansatz techniques; for arbitrary values of $p$ and $q$,
results can be found in \cite{Kim,
Noh_Kim}.

The first part of this article deals with the analysis of the spectrum of $H$
for $q=0$. The Hamiltonian is known to be integrable \cite{HDEV}, but 
because of the
lack of a reference state, the model cannot be treated by a Bethe-Ansatz. 
Therefore numerical methods are applied here. We were interested in 
determining the correspondence between the spatial correlation lengths already
known from previous work described above and the dynamical properties of
the Hamiltonian, i.\ e.\ the time correlation length $\tau$ given by 
the inverse
of the smallest energy gap $E_{G}$  with respect 
to the ground state of the Hamiltonian \cite{Kogut}.
$E_{G}$ is determined for lattice lengths of $2 \leq N \leq 18$ sites 
and then extrapolated to the thermodynamical limit. We find that
$\tau$ remains
finite in the massive phase and diverges algebraically in the massless regime.
The boundary induced phase diagram  known for the stationary state
is reproduced by the spectrum of the Hamiltonian; so  the dynamics of the 
system exhibits the same physical properties as the stationary state. 

The second part of the paper is devoted to the antiferromagnetic version of the 
Hamiltonian $-H$ where $H$ was mentioned above. 
It can no longer
be interpreted as Hamiltonian of a reaction-diffusion sytem, but nevertheless
it describes interesting physical models, e.g. imposing periodic 
boundary conditions it can be viewed as the
logarithmic derivative of the transfer matrix of the six-vertex model 
\cite{Sutherland} and it can be used to describe the 
equilibrium shape 
of a crystal
\cite{van_Beijeren} or a surface growth model \cite{EW}. 
The phase diagram of the antiferromagnetic chain with periodic boundary
conditions has been determined recently \cite{Giuseppe}. A natural
question arises
whether different boundary terms alter the phase diagram in such a way that 
boundary induced phase transitions occur and indeed this phenomenon can be found
again. It manifests itself clearly in the study of the partition function
of the system. Before presenting our examples we give the definition of the
partition functions we are going to use.
For periodical systems the partition function is given by
\be
{\cal Z}= \lim_{N\to\infty}\mbox{tr} z^{\frac{N}{2\pi\xi}(H-e_{\infty}N)}
\ee
for open systems by
\be
{\cal Z}= \lim_{N\to\infty}\mbox{tr} z^{\frac{N}{\pi\xi}(H-f_{\infty}
-e_{\infty}N)}
\ee

Here $\xi$ is a normalization constant, $e_{\infty}$ the bulk free energy
and $f_{\infty}$ the surface free energy.
For a comparison with the results for the asymmetric XXZ chain let us first
repeat some known results for the symmetric XXZ-chain
with different (but not lattice-length dependent)
boundary conditions. It  has been studied extensively by using 
numerical and analytical
methods (Bethe-Ansatz) \cite{xxz1,xxz7}. 
In the regime  $-1 \leq \Delta < 1$ the spectrum is 
massless and  can be described by a representation of the Virasoro algebra with 
central charge  $c=1$. 
The partition function corresponds exactly to the one of
a Coulomb gas with electric charges $n$ and magnetic charges $m$ 
\cite{di_Francesco}
where $n$ corresponds to the total spin of the chain. It can also be
described by a Gaussian field theory with compactification 
radius $R$ \cite{Ginsparg}.
It is given by 
the following expression
(using an appropriate normalization $\xi(\Delta)$:
\be
{\cal Z}=
z^{-\frac{1}{12}}\sum_{m\in Z, n}z^{\frac{1}{2}(\frac{n^2}{R^2}
+R^2m^2)}\Pi^2_V(z)
\label{zper}
\ee
with
\be
\Pi_V(z)=\prod_{i=1}^{\infty}(1-z^i)^{-1}\quad .
\label{eta_function}
\ee
In eq. (\ref{zper}) $n$ takes integer values for an even number of sites
and half-integer values for an odd number of sites.

The spectrum of the chain without any boundary terms is given by a Coulomb gas 
with only electric charges $n$. Here the partition function reads \cite{xxz1}
\be
{\cal Z}=z^{-\frac{1}{24}}\sum_{n\in \frac{Z}{2}}z^{\frac{n^2}{R^2}}\Pi_V(z)
\ee
This expression is valid for the open symmetric XXZ-chain 
as well as for the open 
asymmetric XXZ-chain since there exists a mapping of the symmetric
onto the asymmetric XXZ-chain.

For the periodic asymmetrix XXZ-chain that is normally massive in
the antiferromagnetic region, a massless phase was found 
for $\Delta <1$ \cite{Noh_Kim, Giuseppe}. Here the operator content
contains a term 
which is proportional to the lattice length $N$  so that the partition
function is the one of a modified Coulomb gas. It has been calculated by
Noh and Kim using Bethe-Ansatz calculations \cite{Noh_Kim}.
\be
{\cal Z}
=z^{-\frac{1}{12}}\sum_{m\in Z, n}z^{\frac{1}{2}(\frac
{n^2}{R^2}+ R^2m^2+2imNy)}\Pi^2_V(z)
\ee
The compactification radius $R$ depends on $\Delta$ and $q/p$, $n$
takes integer values for even lattice lengths and half-integer values for
odd lattice lengths. 

These three examples  show that boundary conditions can change the 
spectrum of the asymmetric XXZ-chain drastically. 

The boundary
conditions we analyse in this article (with additional
non-diagonal boundary terms) have not been treated before. Our new results
are obtained by determining the lowest-lying eigenvalues of the 
Hamiltonian up to 18 sites in the general case and 21 sites in the
CP-invariant case ($\alpha = \beta$). As the Hamiltonian does not have 
any symmetries in the
general case 
and is non-hermitian, the diagonalization
requires a large amount of CPU-time.
A version of the deflated Arnoldi algorithm has been applied 
to quantum spin chains for the first time here. 
The Arnoldi algorithm that reduces to the Lanczos algorithm in the case of
hermitian matrices
was already used for the
determination of the spectrum of the Potts model \cite{KAY}. Another
example where a non-hermitian chain was treated numerically can be found
in \cite{Gehlen}.

The analysis of the spectrum for $p>q$ and $\alpha,\beta >0$ suggests
 surprisingly
the partition function of a Coulomb gas that  has only magnetic charges. 
Additionally, the magnetic quantum number $m$ is shifted by an imaginary amount 
$i x$.
\be
{\cal Z}
=z^{-\frac{1}{24}}\sum_{m\in \frac{Z}{2}}z^{R^2(m+i(x+Ny))^2}\Pi_V(z)
\label{zein}
\ee
Here 
 $x=x(\frac{q}{p},\frac{\alpha}{p},\frac{\beta}{p})$ and $y=y(\frac{q}{p})$.
For even lattice lengths $m$ takes half-integer values, for odd lattice lengths
integer values.
The spectrum is
given by  non-unitary representations of the $U(1)$ Kac-Moody algebra.
For $q=0$ the parameters $y$ and $R$ correspond exactly to the ones that
appear for the periodic system \cite{Gwa_Spohn}. The term proportional
to $x$ however
is caused by the boundary terms alone. For $\alpha\beta =0$ the spectrum is
massive, so we can conclude that the boundary terms  give
rise to the conformal invariant structure of the spectrum. 

We also studied a toy model consisting of a simplified version of the
Hamiltonian (\ref{Ham}) that can be diagonalized in terms of free
fermions. The properties of the spectra that have been found numerically
can be found again in the analytical results for the toy model.
Additionally, we gain some insight into the role the different boundary terms
play for the spectrum of the Hamiltonian.

The article is organized as follows: In the first part, we define the total
asymmetric diffusion model with open boundaries and additional injection
and extraction terms. We present our numerical results for
the temporal correlation length in the second part
and compare them to the phase diagram 
of the stationary state and the known expressions for 
spatial correlation lengths. 
The third part deals with the antiferromagnetic version of the Hamiltonian.
From the finite-size scaling behaviour of the lowest-lying energy levels we
conclude the form of the partition function (\ref{zein}). The fourth part
describes analytical results obtained by diagonalization of the toy model
for non-hermitian boundary conditions. Here the
characteristic properties of the Hamiltonian that has been treated 
numerically are reproduced, thus the calculations for the simplified
model give similar results to our numerical analysis of the more complicated 
model.
We close with a discussion of our results. In the appendix tables of the 
numerical results of the second and third part can be found.

\section{Asymmetric diffusion model with boundary terms}

We consider a model defined on a chain with $N$ sites. Each site
can be
either occupied by a particle of species $A$ or empty. For the dynamics,
we consider only processes that involve two neighbouring sites. The 
following processes are allowed:
\be
\begin{array}{lll}
\mbox{diffusion to the right:}&A+0\rightarrow 0+A&\quad
\mbox{with rate } p\\
\mbox{diffusion to the left:}&0+A\rightarrow A+0&\quad
\mbox{with rate } q
\end{array}
\ee
Additionally, we allow processes that occur only at the two edges of
the chain:
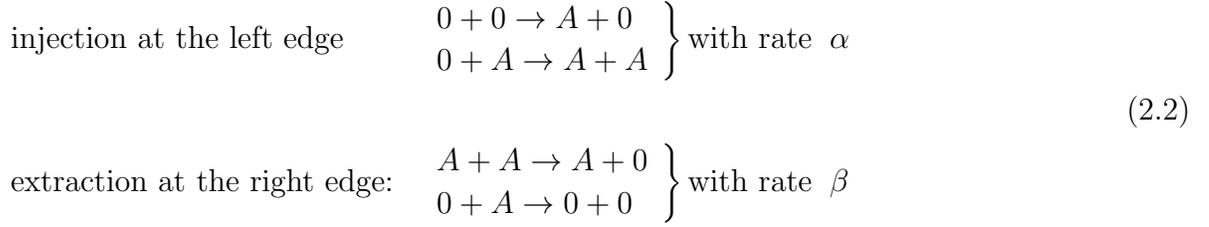
\be
\begin{array}{lll}
\mbox{injection at the left edge}&
\left.
\begin{array}{l}
0+0\rightarrow A+0\\
0+A\rightarrow A+A
\end{array}
\right\}
\mbox{with rate}\;\; \alpha \\[5mm]
\\
\mbox{extraction at the right edge:}&
\left.
\begin{array}{l}
A+A\rightarrow A+0\\
0+A\rightarrow 0+0
\end{array}
\right\}
\mbox{with rate} \;\;\beta \\
\end{array}
\ee

To each site $j$ of the chain we attach a variable $\gamma_j$ that takes the
value $1$ if the site is occupied by a particle and $0$ if the site is empty.
A configuration of the model is described by a set of variables $\{\gamma\}=
\{\gamma_1,
\gamma_2,\ldots, \gamma_N\}$. The probability that a state 
$(\m,\n)$ on two adjacent sites will
change into the state $(\rh,\s)$ after one unit of time is denoted by:
\be
\G^{\m,\n}_{\rh,\s};\;\;\;(\m,\n)\not=(\rh,\s).
\ee
All reactions changing the state $(\m,\n)$ into any other state
are summarized in the rate $\G_{\m,\n}$:
\be
\G_{\m,\n}=
\sum_{\rh,\s =0}^{1}\!\!\!^\prime\,
                   \G^{\m,\n}_{\rh,\s}\;,
\label{nochange}
\ee
where the prime is always used to indicate that in the sum the case
$(\rh,\s)\,=\,(\m,\n)$ is excluded. This definition ensures the conservation
of probabilities.

The dynamics of the system is given by the
time evolution of the probability $P(\{\gamma\},t)$ to find the system in
the configuration $\{\gamma\}$ at time $t$. The time evolution is described
by the master equation \cite{kad}
\begin {eqnarray}
\frac{ \partial}{\partial t}P(\lbrace\gamma\rbrace;t)
     & = & \sum_{k=1}^{L-1}\biggl[
             -\G_{\gamma_{k},\gamma_{k+1}}
                       P(\gamma_{1}, \ldots,\gamma_{L};t)\nn\\*
     &   & +\sum_{\delta_k,\delta_{k+1}=0}^{1}\!^\prime\,
                      \G^{\delta_k,\delta_{k+1}}_{\gamma_{k},\gamma_{k+1}}
                P(\gamma_{1},\ldots,\gamma_{k-1},\delta_k,\delta_{k+1},
                \gamma_{k+2},
               \ldots,\gamma_{L};t)\biggr]\;.
\label {eqn:master}
\end {eqnarray}

The master equation can be mapped on a Euclidean Schr\"{o}dinger equation.
For this purpose, one defines a ket vector  describing the probability 
distribution in a
$2^N$ dimensional vector space $C^{\otimes N}$:
\be
| P(t)> = \sum_{\{\gamma\}}P(\{\gamma\},t)
|\{\gamma\}>\quad .
\ee
The master equation then takes the form
\be
\frac{\partial}{\partial t}| P(t)>=-H | P(t)>
\quad .
\ee
where the Hamiltonian is a sum of nearest neighbor terms
\be
H=\sum_{k=1}^{N-1}H_{k,k+1}
\ee
The $4 \times 4$ matrices $H_{k,k+1}$ are given by the following expression
with respect to the basis of states 
$ | 11 >, | 10 >, | 01 >, | 00 > $ 
\be
H_{k,k+1}=
\left(
\begin{array}{rrrr}
0&0&0&0 \\
0&q&-p&0\\
0&-q&p&0\\
0&0&0&0
\end{array}
\right)
\qquad \mbox{for }k=2,3,...,N-2\quad .
\ee

The boundary terms which arise due to the processes of injection on the first
lattice site and extraction on the last site can be written in this 
formalism as
 \be
H_{1,2}=
\left(
\begin{array}{rrrr}
\alpha&0&0&0 \\
0&q+\alpha&-p&0\\
-\alpha&-q&p&0\\
0&-\alpha&0&0
\end{array}
\right)\quad,\quad
H_{N-1,N}=
\left(
\begin{array}{rrrr}
0&-\beta&0&0 \\
0&q+\beta&-p&0\\
0&-q&p&-\beta\\
0&0&0&\beta
\end{array}
\right).
\ee

The Hamiltonian can also be expressed in terms of Pauli spin matrices
$\sigma^x,\sigma^y$ and $\sigma^z$:
\be
H=-\sum_{j=1}^{L-1}\left[q\sigma_j^-\sigma_{j+1}^+
 +p\sigma_j^+\sigma_{j+1}^-+ \frac{p+q}{4}(\sigma^z_j\sigma^z_{j+1}
-1) + \frac{q-p}{4}(\sigma_{j+1}^z-\sigma_j^z)\right]+B_1+B_N \label{HA}
\ee

with boundary terms
\[
B_1=\quad\frac{\alpha}{2}(\sigma_1^z-2\sigma_1^- 
+1)\quad ,
\]
\be
B_N=-\frac{\beta}{2}(\sigma_N^z +2\sigma_N^+
-1)\quad .
\ee

The spectrum of $H$ stays invariant if $\alpha$ and $\beta$ are permuted. 
This permutation can be obtained on the level of the Hamiltonian
by applying the similarity transformation
$\sigma^{\pm} \rightarrow \sigma^{\mp}, \sigma^{z} \rightarrow -\sigma^{z}$
and a simultanous reflection of the system. Physically one can understand this 
property by denoting that particle-hole symmetry holds: The model
can be seen as describing particles moving in one direction or holes 
moving in the opposite  direction. If $\alpha = \beta$, the Hamiltonian 
is invariant under this transformation ($CP$-symmetry).

The Hamiltonian can also be mapped by a unitary transformation $U$ 
onto the symmetric XXZ-chain with different boundary terms. 
This correspondence
will prove very helpful in the sequel and therefore it will be given explicitly
here.
If we take 
\be
U=\prod_{j=1}^{N}U_j\quad,\quad U_j=I_1\otimes\cdots I_{j-1}\otimes
\left(
\begin{array}{ll}
        1 & 0 \\
        0 &\Lambda \Q^{j-1}
\end{array}
 \right)\otimes I_{j+1}\otimes\cdots\otimes I_N \label{uni}
\ee
with $\Q=\sqrt{\frac{q}{p}}$ and arbitrary $\Lambda$
the transformed Hamiltonian $H^{\prime} = U H U^{-1}$ is:
\[
H'=-\frac{\sqrt{pq}}{2}\sum_{j=1}^{N-1}\left[ \sigma_j^x\sigma_{j+1}^x
 +\sigma_j^y\sigma_{j+1}^y
+ \frac{\Q+\Q^{-1}}{2}(\sigma^z_j\sigma^z_{j+1}-1)\right.
\]
\be
\qquad \left. + \frac{\Q-\Q^{-1}}{2}(\sigma_{j+1}^z-\sigma_j^z)\right]+B'_1+B'_N
\quad.
\label{H'}
\ee
This is the $U_{\Q} SU(2)$ invariant Hamiltonian \cite{Pasquier_Saleur}
with additional 
length-dependent boundary terms:
\[
B'_1=\quad\frac{\alpha}{2}(\sigma_1^z-2\Q^{\frac{1-N}{2}}
\sigma_1^- 
+1)
\quad ,
\]
\be
B'_N=-\frac{\beta}{2}(\sigma_N^z
+2\Q^{\frac{1-N}{2}}\sigma_N^+
-1)\quad . \label{rand'}
\ee
Although this Hamiltonian is known to be integrable \cite{HDEV}, 
it cannot be solved
by the Bethe-Ansatz because one cannot construct a reference state. 
The non-hermitian boundary terms induce completely different properties of
the Hamiltonian $H$ in comparison to $H'$ without $B_1'$ and $B_2'$. The reason
for this
can be understood by considering the matrix form of $H$. As the bulk terms
(without $B_1^{\prime}$ and $B_2^{\prime}$) commute with the total spin $S^z = 
\sum_{j=1}^{N} \sigma^z$, these bulk terms can be written as a matrix 
in block-diagonal
form, where each block operates in a sector with fixed total spin. The boundary
terms change the total spin by $\pm 1$, so they will appear in blocks below
or above the diagonal. Only in the case where $\alpha \neq 0$ {\bf and}
$\beta \neq 0$, they will give contributions to the spectrum. In all other
cases the spectrum of $H$ is the massive spectrum of the  $U_{\Q} 
SU(2)$-invariant XXZ-chain.

\section{Results for the total asymmetric diffusion model with boundary terms}
\subsection{Analytical results for the stationary state}
For future reference, let us first present some of the results achieved 
previously for the total asymmetric diffusion model. Up to now,
all results are  known for the stationary state.
In \cite{DEHP} the phase diagram for the current and the spatial profile
of the concentration have been determined as functions of $\alpha$ and $\beta$.
The current through the bond $j$ for the configuration ${\gamma}$ is
defined as:
\be
j_k=\sum_{\{\gamma\}}\gamma_k(1-\gamma_{k+1}) P(\{\gamma\},t)\quad .
\label{strom}
\ee

The phase diagram is given in figure 1. There are three main phases, 
characterized by the density of particles: The high-density phase $A$, the 
low-density phase $B$ and the maximal current phase $C$. Phases $A$ and $B$ are
further subdivided into $A_1, A_2$ and $B_1,B_2$. Phases $A$ and $B$ are
separated by a line which is called 'coexistence line'. In the thermodynamic 
limit, the currents in the three phases are given by:
\[
A\quad :\quad j=\beta(1-\beta)
\]
\be
B\quad :\quad j=\alpha(1-\alpha)\quad .
\ee
\[
C\quad :\quad j=\frac{1}{4} 
\]
The density profile of the concentration on the stationary state obtained in
\cite{DEHP} allows to read off the spatial correlation length $\xi$ in the 
different phases. In the pase $A$, $\xi$ is 
defined by 
\be
<n_k> = \mbox{const.} + \mbox{const.} \exp(-k/\xi)\;\;;
\ee
in the phase $B$ it is given by
\be
<n_k> = \mbox{const.} + \mbox{const.} \exp(\frac{k-N-1}{\xi})\;\;.
\ee
In the different phases the expressions for the correlation length are: 
\ba
A_1:\xi & =& \Bigl(\ln(\frac{\alpha(1-\alpha)}{\beta (1-\beta)})\Bigr)^{-1}\nn\\
A_2:  \xi & = & - \ln (4 \beta (1-\beta))\nn\\
B_1: \xi & = & \Bigr[\ln (\frac{\beta(1-\beta)}{\alpha (1-\alpha)})\Bigl]^{-1}
\nn\\
B_2: \xi & = & - \ln (4\alpha (1-\alpha))
\label{corr_length1}
\ea
In the phase $C$ and on the coexistence line the one-point function 
$<n_k>$ shows
an algebraic behaviour:
\ba
C: <n_{N-k}>& = &\frac{1}{2}-\frac{1-\delta_{\beta,\half}}
{2 \sqrt{\pi}} k^{-\frac{1}{2}}
\nn\\
\mbox{coexistence line}: <n_k> & = & \alpha + k \frac{1-2 \alpha}{N}
\label{corr_length2}
\ea

\subsection{Numerical results for the time correlation length}
In this chapter we compare the phase structure of the
stationary state to the spectrum of the Hamiltonian.
We investigate the gap between the ground state  and the first excited state of 
the spin chain. This energy gap $E_G$ allows us to read off the time 
correlation length $\tau$ directly:
\be
\tau^{-1} \simeq E_G\;\;\;.
\ee
We determine the lowest lying excitation energies for lattice lengths
$2 \leq N \leq 18$ by diagonalizing the Hamiltonian numerically,
using a version of the deflated Arnoldi algorithm \cite{Arnoldi}
that reduces to the Lancos algorithm in the case of hermitian matrices.
The eigenvalues were then extrapolated with the help of
 the BST-algorithm \cite{BST}.
$\alpha$ and $\beta$ have been varied in steps of $0.1$ between $0.1$ and $1$.
The following form of the energy gap has been found in the different regions of
 the phase diagram. In the phases $A_1$ and $B_1$, $E_G$ is a function of 
$\alpha$
and $\beta$, while in phase $B_2$, $E_G$ depends only on $\alpha$ and in phase 
$A_2$ only on $\beta$. In phase $C$ and on the coexistence 
line the system is massless:
\begin{eqnarray}
A_1\quad :\quad & E_G = m(\alpha,\beta)\nonumber \\
A_2\quad :\quad & E_G = m(\beta)\nonumber \\
B_1\quad :\quad & E_G = m(\alpha,\beta)\nonumber \\
B_2\quad :\quad & E_G = m(\alpha)\\
C  \quad :\quad & E_G \sim N^{-\frac{3}{2}}\nonumber \\
\alpha=\beta<\frac{1}{2} \quad :\quad & E_G \sim N^{-2}\nonumber
\end{eqnarray}
Here $m$ denotes the mass of the spectrum. 

Table I shows extrapolants for the energy gaps. One sees clearly that in the
phase $B_1$ the energy gap depends on both $\alpha$
and $\beta$ while it depends only on $\alpha$ in the phase $B_2$. We
did not give any extrapolants for the phases $A_1$ and $A_2$ since 
the spectrum is symmetric under permutation of $\alpha$ and $\beta$
as described in section 2. Table I also shows that the
mass gap vanishes in the phase C and on the coexistence line. In Table II
extrapolants for the exponent $\frac{3}{2}$ in the phase $C$ and
for the exponent $2$ on the coexistence line can be found.  

Comparing these results with (\ref{corr_length1}-\ref{corr_length2}), we can
see that the temporal correlation length shows the same behaviour as
the spatial correlation length in  
the different phases: in the phase $C$ and on the coexistence line $\tau$ 
diverges 
algebraically while it remains finite in the phases $A$ and $B$. 
The exponent $\frac{3}{2}$ has also been found for the total asymmetric 
diffusion model with
periodic boundary conditions \cite{Gwa_Spohn}. The exponent $2$
was found for the symmetric XXZ chain with periodic boundary conditions
at the point $\Delta=1$. The periodic 
model can be mapped onto a model for surface growth \cite{Gwa_Spohn, Kim, 
Neergard_Nijs}. This mapping can be formulated analogously for the
open chain with additional boundary terms that is treated here. 
Then, in the language of 
growth models, the exponent $\frac{3}{2}$ describes KPZ-type growth \cite{KPZ}
while the exponent $2$ corresponds to Edwards-Wilkinson growth 
behaviour \cite{EW}.  
In phases $A_2$ and $B_2$ $\tau$ depends only on $\beta$ or $\alpha$
respectively. However, the numerical values for the temporal correlation
length are different from the ones for the spatial correlation length.
The fact that the temporal evolution of the system reflects itself in
the stationary state (for $t \rightarrow \infty$) is not yet understood.
This is not true in general, as there are examples for a different 
behaviour of the spatial and the temporal correlation length \cite{bibel}.
However, in this case it is a striking and unexpected feature that shows
how strong boundary conditions change the properties of non-hermitian
quantum chains. 

\section{The antiferromagnetic chain}
\label{antiferro}
In the following, we will treat the antiferromagnetic Hamiltonian which can be
obtained from the ferromagnetic one (that has been treated before) 
by applying a similarity transformation. As has been shown in chapter 2 
the ferromagnetic chain can be connected to a XXZ-chain with anisotropy
$\Delta = \frac{p+q}{2} \geq 1$. The antiferromagnetic chain corresponds
to $\Delta \leq 1$
and can be written as
\[
H''=-\sum_{j=1}^{L-1}\left[q\sigma_j^-\sigma_{j+1}^+
 +p\sigma_j^+\sigma_{j+1}^-- \frac{p+q}{4}(\sigma^z_j\sigma^z_{j+1}
-1) - \frac{q-p}{4}(\sigma_{j+1}^z-\sigma_j^z)\right]+B''_1+B''_N 
\]
with boundary terms
\[
B''_1=\quad -\frac{\alpha}{2}(\sigma_1^z+2\sigma_1^-
+1)
\quad ,
\]
\be
B''_N=\frac{\beta}{2}(\sigma_N^z + (-1)^N 2\sigma_N^+
-1)\quad .
\ee

Due to the minus sign in the transformation rule, the low lying excitations
of the ferromagnetic chain correspond to the highest states of the 
antiferromagnetic chain and vice versa.

The numerical studies concentrate on the case $q=0$ and $p=1$ with $\alpha,\beta
\geq 0$. For this choice of parameters the extrapolated values converge very 
well.

The analysis of the spectrum reveals that it is conformal 
invariant. We studied the finite-size scaling behaviour of the model 
in order to determine the operator content of the underlying conformal
field theory. For conformal invariant systems, the ground state is supposed
to take the following form for finite lattices and for
periodic boundary conditions \cite{BCN,Affleck}
\be
\frac{E_0(N)}{N}=e_{\infty}-\frac{\pi\xi c}{6 N^2} - o(N^{-2})\quad 
\label{e0p}
\ee
The excited state $E_r$ satisfies \cite{Cardy} 
\be
{\cal E}_r=\lim_{N\to\infty}\frac{N}{2\pi\xi}(E_r(N)-E_0(N))= (\Delta +r)
+(\bar{\Delta} +\bar{r})\quad;
\ee
for open boundary conditions, one gets an additional surface term $f_{\infty}$
in the ground state energy \cite{Cardy2}:
\be
\frac{E_0(N)}{N}=e_{\infty}+\frac{f_{\infty}}{N}-\frac{\pi\xi c}{24 N^2} 
- o(N^{-2})\quad .
\label{cc}
\ee
The energy gaps scale here as
\be
{\cal E}_r=\lim_{N\to\infty}\frac{N}{\pi\xi}(E_r(N)-E_0(N))= (\Delta +r)
\quad;
\ee
Here $\xi$ is a normalization constant, $c$ the central charge of the Virasoro
algebra. 

In our case, the spectrum had to be treated separately for even and odd lattice 
lengths. Since the ground state was calculated for an odd number of sites,
we used interpolated values from odd lattice lengths for the values of
the ground state for even lattice lengths.
While the ground state energy is real for all lattice lengths, most
of the excited states have a non-vanishing imaginary part. This is a  feature
that did already appear for the periodic asymmetric XXZ chain \cite{Noh_Kim}
and in the calculation of the operator content of the
five vertex model defined on an anisotropic lattice
\cite{Kim_Pearce}. 

We considered the imaginary part of the energy gap
\be
{\cal I} = \lim_{N\rightarrow\infty}\frac{1}{\pi\xi}\mbox{Im} (E(N))\quad,
\ee
the correction to the real part 
\be
\mbox{Re}({\cal E}) = \lim_{N\rightarrow\infty}\frac{N}{\pi\xi}
\mbox{Re}(E(N)-E_0(N)))\quad
\ee
and the correction to the imaginary part
\be
\mbox{Im}({\cal E}) = \lim_{N\rightarrow\infty}\frac{N}{\pi\xi}
\mbox{Im}(E(N)-{\cal I})\quad
\ee

The data reveals the following finite-size scaling behaviour for the normalized
eigenvalues:
\be
\frac{N}{\pi\xi}(E^m_r-E'_0)=R^2(m+i(x+yN))^2+r\quad,
\quad r\in N\quad,\label{op}
\ee
and the ground state behaves as
\be
E'_0=E_0+\pi\xi(\frac{R^2}{N}(x+yN)^2\quad . \label{E'}
\ee
Here $m$ is a quantized number and takes for odd lattice lengths the integer 
values $m = 0, \pm 1, \pm 2,\cdots$, for even lattice lengths 
half-integer values
$m = \pm \frac{1}{2},\pm \frac{3}{2}, \ldots$. The precise relation between
the extrapolants and the constants $R,x$ and $y$ is given by:
\[
{\cal I}^m_r = 2R^2my,
\]
\be
\mbox{Re}({\cal E}^m_r) = R^2m^2 + r, \label{extrap}
\ee
\[
\mbox{Im}({\cal E}^m_r) = 2R^2mx.
\]
Data for $r=0$ is shown in table III.
As $H$ is real, we find for each complex eigenvalue $E$ also the complex
conjugate $E^{*}$. Therefore all tables only show data for $m \geq 0$. Data
for $r \geq 0$ can be found in table IV for $m=0$ and $m=\frac{1}{2}$. 
The corresponding data for $m=1, \frac{3}{2},2,\frac{5}{2}$ and $3$ 
has also been obtained.

The next constant that has to be determined is the conformal 
charge $c^{\prime}$. The ground state energy in the thermodynamic limit,
$e_{\infty}$, is already known from the periodical system. The surface
energy $f_{\infty}$ can be obtained by extrapolation of
\be
f_{\infty}=\lim_{N\to\infty}(E_0(N)-Ne_{\infty})\quad .
\ee
The results are given in table V.
Now $c$ can be obtained via
\be
c=\lim_{N\to\infty}\frac{24N}{\pi\xi}(E_0(N)-Ne_{\infty}-f_{\infty})\quad.
\ee
The numerical values for $c$ are not constant for different values of $\alpha$
and $\beta$ (see table VI). 
However, we can shift all energy gaps, independently of the 
sector by a constant term  $24 R^2 x(\alpha,\beta)^2$ that depends on 
$\alpha$ and $\beta$ (table VII).  Absorbing this shift into the ground state, 
one can
define a new central charge $c^{\prime}$ 
\be
c'=c-24R^2x^2\quad .
\ee
which is indeed constant within the numerical errors.
The numerical values show an excellent agreement with $c^{\prime}=1$ (table VI).

The term $\pi \xi I^m_r$, the imaginary part of the energy gaps, takes
(independently of $\alpha$ and $\beta$) for odd lattice lengths multiples
of the same constant $1.8854....$ that already appeared in the calculation
of Gwa and Spohn \cite{Gwa_Spohn} for the periodic system as imaginary part
of the smallest energy gap. Using Bethe Ansatz calculations 
for the first and second
smallest eigenvalue in the sector with spin $0$ they obtained the 
following result 
\be
e_{\infty}=0.690140115 .\label{fbe}
\ee
\be
(E_1-E_0)_{per}=6.5776787 N^{-1}+i 1.885456427 . \label{gwap}
\ee

This result can be used to obtain estimates for the parameters $R$ and $y$,
if one assumes that also the spectrum of the periodic chain is characterized
by these parameters, and the finite-size scaling of the lowest lying state
with spin $0$
is given by \cite{Noh_Kim}
\be
\frac{N}{2\pi\xi}(E_1-E_0)_{per}=\frac{1}{2}R^2+iR^2yN \;\;.
\ee

The estimates that have been obtained using this assumption can now be compared
to the numerical results. Tables III and IV show a comparison between the 
values obtained
from the analytical calculation and the numerical results. The normalization 
constant $\xi$ has been taken from numerical Bethe Ansatz calculations for
the periodical system for up to $80$ sites \cite{BirSil}: 
\be
\xi=1.64784392694623 \label{xi}\quad .
\ee
The data shows that indeed the 
periodic system is characterized by the same parameters as the open system
with additional boundary terms. However, the full operator content of
the two chains is  different. This effect comes from
the different boundary terms only. 

The degeneracies of the energy levels for $r = \mbox{const.}$ are described by
the characterfunction of a $U(1)$ Kac-Moody algebra \cite{GOL}. This
confirms the above result $c'=1$. The Kac-Moody algebra is defined
by its commutation relations 
\be
[T_m,T_n]=m\delta_{m+n,0}\quad ,\quad m,n\in Z\quad .
\label{km}
\ee
The character function is given by
\be
\chi_{\Delta,q}(z,y)=\mbox{tr}(z^{L_0}y^{T_0})=z^{\frac{q^2}{2}}
\Pi_V(z)y^q\quad . \label{kaccha}
\ee
where $L_0$ is a generator of the Virasoro algebra with conformal weight
$c=1$ that can be canonically obtained from the $U(1)$
Kac-Moody algebra using the
Sugawara construction \cite{COR}. Here $q$ is the eigenvalue of $T_0$, and the
highest weight representation is $[\Delta,q]$. A shift in the algebra 
characterized by a parameter $\varphi$
\be
\tilde{T}_m=T_m+\varphi\delta_{m,0}
\ee
does not change the commutation relations above but leads to
a representation with highest weights
\be
\tilde{\Delta} = \frac{1}{2} \tilde{q}^2 \;\;\;\mbox{where} \;\;\;\;\;
\tilde{q} = q +\varphi 
\ee
If $\varphi$ is chosen to be complex, one obtains a representation of the
Kac-Moody algebra with negative conformal dimensions. In this case however
$\tilde{T^{+}_0} \neq \tilde{T_0}$ and the representation is not unitary.  

In our case, the part of the energy corrections that is independent of $N$
can be described by a non-unitary representation of a shifted
Kac-Moody algebra. The parameter $x$ depends on the boundary terms.
The shift $\varphi$ of the Kac-Moody algebra is given by $\varphi=\sqrt{2}
i(x+Ny)$. 

The results for the antiferromagnetic chain can be summarized in the partition
function
\be
{\cal Z}
=z^{-\frac{1}{24}}\sum_{m\in \frac{Z}{2}}z^{R^2(m+i(x+Ny))^2}\Pi_V(z)
\ee
The operator content of this model corresponds to the one of a Coulomb
gas with only magnetic charges and an additional term that depends on the 
lattice length $N$.
All calculations for $q \neq 0 \neq p$ for $p>q$ show
similar results but reveal that in this case $x$ and $y$ are functions of the
ratio $q/p$. 

The case $q>p$ has not been studied systematically. Here the spectra a purely
real, but the convergence of the extrapolations is too bad to obtain 
precise estimates (the same holds for the case $p=q$, i.e. 
the symmetric $XXZ$ chain
with additional boundaries).  

\section{Toy model}
In this chapter we study analytically
a modified version of the Hamiltonian treated before.
The results for different boundary conditions
show that the main new features in comparison to the hermitian
chain can already be found in this toy model. This model allows to
understand which term in the Hamiltonian is responsible for the new
and unexpected contributions to the spectrum.
 
We will concentrate on
\be
H=\sum_{j=1}^{N-1}\sigma^+_j\sigma^-_{j+1} +\alpha\sigma_1^-+\beta\sigma_N^+
\ee
For $\alpha=0$ or $\beta=0$ the spectrum consists only of the $N$ times
degenerate eigenvalue zero. So the whole structure of the spectrum is caused
by both boundary terms together. 

This Hamiltonian can be diagonalized in terms of free fermions. Moreover
$H(\alpha,\beta)$ can be transformed into $H(-\alpha,-\beta)$
by applying the transformation $\sigma^{\pm}\to -\sigma^{\pm}$. For
$\alpha\beta>0$ the characteristic properties of the spectra 
described previously 
are reproduced.

In order to write $H$ in terms of free fermions, we have to obtain a bilinear 
expression in $\sigma$-matrices so that standard fermionisation techniques
can be applied \cite{LSM}. 
Technically, this can be achieved by appending one lattice site
at each end of the chain, site $0$ and site $N+1$ \cite{Peschel}. 
The Hamiltonian then reads
\be
H'=\sum_{j=1}^{N-1}\sigma^+_j\sigma^-_{j+1}
+\alpha\sigma^x_0\sigma_1^-
+\beta\sigma_N^+\sigma^x_{N+1}
\ee
As $\sigma_0^x$ and $\sigma_{N+1}^x$ commute
with $H'$ thus being constants of motion, the spectrum of $H'$ 
decomposes into four sectors
$(++, +-, -+,--)$ corresponding to the eigenvalues $\pm 1$ of $\sigma_0^x$ and
$\sigma_{N+1}^x$. 
The eigenvectors of the extended Hamiltonain include those of $H$.
Therefore we can obtain the eigenvectors of the original problem
by projecting onto the $(++)$-sector.

Defining new operators $\tau_j^+$ and $\tau_j^-$ \cite{Haye} by
\be
\tau_j^+=(\prod_{i<j}\sigma^z_i)\sigma_j^x ,\qquad
\tau_j^-=(\prod_{i<j}\sigma^z_i)\sigma_j^y  \qquad (j=0,...,N+1)
\ee
that obey the anticommutation relations of a Clifford-algebra
\be
\{ \tau_i^{\mu},\tau_j^{\nu}\}=2\delta_{i,j}^{\mu,\nu}
\qquad (i,j=0,..,N+1; \mu,\nu=\pm 1).
\ee
one can rewrite $H'$ as a bilinear expression in $\tau_j^+$ and $\tau_j^-$
\be
H'=-\sum_{\mu,\nu=\pm 1}\sum_{j=1}^{N-1} A_j^{\mu,\nu}\tau^{\mu}_j
\tau^{\nu}_{j+1} + B^{\mu,\nu}\tau^{\mu}_0\tau^{\nu}_{1}+
B^{\mu,\nu}\tau^{\mu}_N\tau^{\nu}_{N+1}
\ee

with
\be
B=\frac{\alpha}{2}
\left(
\begin{array}{ll}
        1 & i \\
        0 & 0
\end{array}
 \right),\quad
A=\frac{1}{4}
\left(
\begin{array}{ll}
        1 & i \\
        -i & 1
\end{array}\right)\quad(0<j<N)\quad,\\
C=\frac{\beta}{2}
\left(
\begin{array}{ll}
        0 & i \\
        0 & 1
\end{array}
 \right).
\label{A}
\ee

Here we chose as a basis
\be
A=
\left(
\begin{array}{ll}
        A^{--} & A^{-+} \\
        A^{+-} & A^{++}
\end{array}
 \right)\quad .
\ee
Then a second linear transformation
\be
T_n^{\gamma}=\sum_{j=0}^N\sum_{\mu=\pm 1} (\psi^{\gamma}_n)^{\mu}_j 
\tau^{\mu}\qquad
(\gamma=\pm 1)
\ee
with
\be
\{ T_i^{\mu},T_j^{\nu}\}=2\delta_{i,j}^{\mu,\nu}
\qquad (i,j=0,..,N+1; \mu,\nu=\pm  1).
\label{Tcliff}
\ee
yields
\be
H'=\sum_{n=0}^{N+1}\Lambda_n\, iT^-_n T^+_n \label{hfer}
\ee
The eigenvalues $\Lambda_n$ and the vectors
$\psi^{\gamma}_n=((\psi^{\gamma}_n)^-_0,(\psi^{\gamma}_n)^+_0,...,
(\psi^{\gamma}_n)^-_{N+1},
(\psi^{\gamma}_n)^+_{N+1})$
are given by the solution of the eigenvalue problem
\be
M(\psi_n^+\pm i\psi_n^-)=\mp \Lambda_n(\psi_n^+\pm i\psi_n^-)\quad .
\ee

where $M$ is a $(2N+4)\times(2N+4)$ matrix which is given by
\be
M=
\left(
\begin{array}{cccccc}
0 &B\\
-B^T & 0 & A\\
&-A^T&0&A\\
&&...&...&...\\
&&&-A^T&0&C\\
&&&&-C^T&0\\
\end{array}
\right)\quad ,
\ee

The digonalization of $M$ leads to one eigenvalue $0$ and the following equation
determining the other eigenvalues:
\be
(2\Lambda)^{2N+2}=4\alpha^2\beta^2(-1)^N\quad .
\ee
leading to the solutions
\be
\Lambda_0=0,\quad 2\Lambda_n=(2\alpha\beta)^{\frac{1}{N+1}}
(\sin(k)-i\cos(k)) \label{Disp}
\ee
with
\be
k=\frac{2n-1}{2N+2}\pi \quad (0<n\leq N+1)\quad .
\ee
Because the operators $iT_n^- T_n^+$ always have eigenvalues $\pm 1$
it is sufficient to determine the eigenvalues $\Lambda_n$ with a positive real
part. The energy gaps of the spectrum are given by $2\Lambda_n$.
For the calculation of the exact ground state, one has to sum all
levels with negative energy (so that the fermi see is filled), which leads to
\be
E_0=-\sum_{n=1}^{N+1}\Lambda_n=-\frac{(2\alpha\beta)^{\frac{1}{N+1}}}
{2\sin(\frac{\pi}{2N+2})}\qquad .
\ee

\subsection{Scaling of the energy levels}
For large $N$ the ground state reads
\be
-E_0\approx\frac{N}{\pi}+\frac{1+\ln(2\alpha\beta)}{\pi}+
\frac{\pi}{24N}\left(1+\frac{12\ln^2(2\alpha\beta)}{\pi^2}\right) + 
o(N^{-2})\qquad .
\ee
For the study of the low-lying excitations one has to develop the 
expression (\ref{Disp}) around the minimum of the dispersion relation
which is obtained for momentum $\pm \frac{\pi}{2}$.
For $0<n\ll N$ one obtains:
\be
2\Lambda_n\qquad\approx \frac{2n-1}{2}\frac{\pi}{N}-i\left(1+\frac{\ln(2
\alpha\beta)}{N}\right)
\qquad\mbox{for}
\qquad k<\frac{\pi}{2},
\label{kk}
\ee
\be
2\Lambda_{N+2-n}\approx \frac{2n-1}{2}\frac{\pi}{N}+i\left(1+\frac
{\ln(2\alpha\beta)}{N}\right)
\qquad\mbox{for}
\qquad k>\frac{\pi}{2}.
\label{kg}
\ee
Up to now we always treated the chain where two additional sites, $0$ and
$N+1$ have been added to the starting Hamiltonian. We obtain the spectrum
of the Hamiltonian without these sites by projection onto the sector where
the $\sigma$-matrices acting on the additional sites have eigenvalue $1$.
It can be shown that the spectrum of $H$ with respect to the ground state
$E_0$ for large $N$ is obtained for even $N$ by combining only
an odd number of fermions with energies $2 \Lambda_n$ and for
odd lattice lengths by combining an even number of fermions. The ground
state of the system is found to be in the sector with odd lattice lengths.

For the normalization constant, one reads off directly $\xi=1$. 

We turn now to the determination of the partition function because this
can be compared directly to previous results. 
Writing $\tilde{H}=\frac{N}{\pi}(H-E_0)$
and using the triple product identity \cite{Ginsparg} we obtain
\begin{eqnarray}
\mbox{tr} z^{\tilde{H}}&=\prod_{n=1}^{\infty}(1+z^{n-\frac{1}{2}}z^{ia})
(1+z^{n-\frac{1}{2}}z^{-ia})\nonumber\\
&=\sum_{m\in{\frac{Z}{2}}}^{\infty}z^{2(m+i\frac{a}{2})^2+\frac{a^2}{2}}\prod_{n
=1}^{\infty}
(1-z^n)^{-1}
\end{eqnarray}
with
\be
\frac{a}{2}=x+yN,\quad x=\frac{\ln(2\alpha\beta)}{2\pi},\quad y=\frac{1}{2\pi}
\qquad .\label{fermiy}
\ee
Here we obtain $m\in Z$ for odd lattice lengths and $m\in Z+\frac{1}{2}$ for
even lattice lengths. Absorbing the term $\frac{a^2}{2}$ into the ground state
energy, we get the following expression:
\be
E'_0(N)=-\frac{N}{2\pi}-\frac{1}{\pi}-\frac{\pi}{24N} + o(N^{-2})\quad .
\ee
This corresponds to a system with conformal charge $c=1$. 
For the partition function we obtain
\be
{\cal Z}=z^{\frac{1}{24}}\sum_{m\in{\frac{Z}{2}}}^{\infty}z^
{2(m+i(x+Ny))^2}\prod_{n=1}^{\infty} (1-z^n)^{-1}\quad .\label{zoffen}
\ee

This is exactly the same result we found for the more complicated model
with $R^2=2$.

Comparing the partition function for the toy model with the ones for the 
open and for the periodic XXZ-chain given in the introduction,
we get some insight into the nature of the different new terms in the partition
function. For asymmetric XXZ-chains without periodic boundary conditions,
we get the spectrum of a Coulomb gas with one kind of charges only. In the
case of the completely open chain we get electrical charges, in the case of
non-diagonal, non-hermitian chains we get only magnetic charges. These
boundary conditions also give rise to an imaginary shift $ix$ in the magnetic
charges that also appears for the periodic chain with a real twist
\cite{xxz1}. In our case $x$ depends
on $\alpha, \beta$ and the ratio $p/q$. 
The lattice length-dependent term $yN$ appears for all non-hermitian
chains we mentioned in this article except for the completely open
chain which is equivalent to the hermitian symmetric open XXZ-chain.

\section{Conclusions}
In this article, the spectrum of the asymmetric XXZ-chain with
non-hermitian boundary terms has been studied. Numerical methods have been 
applied for the determination of the eigenvalues. Analytically, we studied a toy
model that reproduces the characteristic properties of the full Hamiltonian.
In two different cases, we concluded that the boundary terms are
responsible for the phase transitions of the Hamiltonian.

We studied first the ferromagnetic chain with $\alpha,\beta >0$ that 
describes the time evolution of a reaction-diffusion system with
asymmetric diffusion in the bulk and additional injection and extraction
terms at the boundaries. For the completely asymmetric chain the behaviour 
of the smallest energy gap that
corresponds directly to the inverse temporal correlation length shows 
the same behaviour as the spatial correlation length in the stationary state
determined by \cite{DEHP}. In phases $A$ and $B$ where the spatial correlation
length stays finite, i.e. the concentration of particles shows an
exponential decay in the spatial direction, we find a massive spectrum where
the mass depends on the same parameters as the spatial correlation length in
the different phases  $A_1,A_2,B_1$ and $B_2$. In this way the subdivision
of the phases $A$ and $B$ is also valid for the dynamical properties 
of $H$. However, the numerical values of the spatial and the temporal 
correlation
lengths do not coincide. 
The phase $C$, where the spatial correlation length of the stationary state 
decays algebraically, exhibits a decay of the temporal correlation length
with an exponent $\frac{3}{2}$. This exponent has already been
identified for the model with periodic boundary conditions.
On the coexistene line, we find a decay with an exponent $2$. The deep
reason why the phase diagram of the stationary state is reflected in
the time-evolution of the system is not yet understood.

The antiferromagentic chain for $\Delta <1$ shows a completely different 
behaviour. For $\alpha =0$ or $\beta  =0$ the spectrum corresponds to that
of a XXZ-chain with additional $\sigma^z$-terms at the boundaries and therefore
is massive. Only when both $\alpha$ and $\beta$ are non-zero, the spectrum is
massless and can be  described by a representation of a non-unitary
$U(1)$ Kac-Moody algebra. However, one term proportional to the lattice
lenght arises in the finite-size scaling spectrum.   
The result can be
summarized in the partition function of a modified Coulomb gas 
with only magnetic charges:
\be
{\cal Z}
=z^{-\frac{1}{24}}\sum_{m\in \frac{Z}{2}}z^{R^2(m+i(x+Ny))^2}\Pi_V(z)
\ee
The parameters $y$ and $R$ also appear for periodical boundary conditions
while $x$ is induced by the boundary terms.
The analysis of the toy model that was diagonalized in
terms of free fermions reproduces the same structure of the spectrum. 
The length-dependent term seems to be a common property of the
anisotropy of the spin chain \cite{Kim_Pearce}
and also appeared for the periodic
chain \cite{Noh_Kim}. The imaginary shift of the magnetic charge 
is an effect of the non hermitian and non-
diagonal boundary terms. The influence of hermitian boundary terms 
and asymmetric interactions in the bulk will be the subject of a future 
publication \cite{future}. 

The question arises now how the ordinary Coulomb gas model has to be modified
so that imaginary and lattice-length dependent contributions to the 
operator content arise. Is it still possible to find
a field theory that reproduces the  spectrum 
of the modified Coulomb gas? A first step in this direction
could be the calculation of the correlation functions that normally, for
conformal invariant systems, exhibit critical exponents which are
directly related to the eigenvalues of the corresponding Hamiltonian.
Wether this relation is still valid in the case of non-hermitian chains still 
has to be clarified.

The new and unexpectedly interesting structure that appeared in the two
examples treated in this article suggests further research in the field
of non-hermitian Hamiltonians.

\section*{Acknowledgements}
We wish to thank Prof.\ V.\ Rittenberg for suggesting this interesting
problem and for enlightening discussions and constant support. We are very
grateful to Prof.\ D.\ Kim for valuable discussions and suggestions. 
We wish to thank the whole group for constant help and discussions.

\newpage
\listoffigures
\listoftables

\unitlength2.5cm
\begin{figure}
\begin{center}
\begin{picture}(3,4)
\put(0,.5){\vector(0,1){3.2}}
\put(0,.5){\vector(1,0){3.2}}
\put(1.5,2){\line(0,1){1.5}}
\put(1.5,2){\line(1,0){1.5}}
\put(0,2){\dashbox{.1}(1.5,0){}}
\put(1.5,.5){\dashbox{.1}(0,1.5){}}
\put(0,.5){\line(1,1){1.5}}
\put(-.2,.3){\large\bf 0}
\put(1.4,.3){\large\bf 0.5}
\put(-.5,2){\large\bf 0.5}
\put(2.2,2.7){\huge\bf C}
\put(2.2,1){{\huge\bf A}2}
\put(1,1){{\huge\bf A}1}
\put(.5,2.7){{\huge\bf B}2}
\put(.5,1.5){{\huge\bf B}1}
\put(3,.3){\large\boldmath$\alpha$\unboldmath}
\put(-.4,3.6){\large\boldmath$\beta$\unboldmath}
\end{picture}
\caption
{Phase diagram for the total asymmetric diffusion model}
\end{center}
\end{figure}
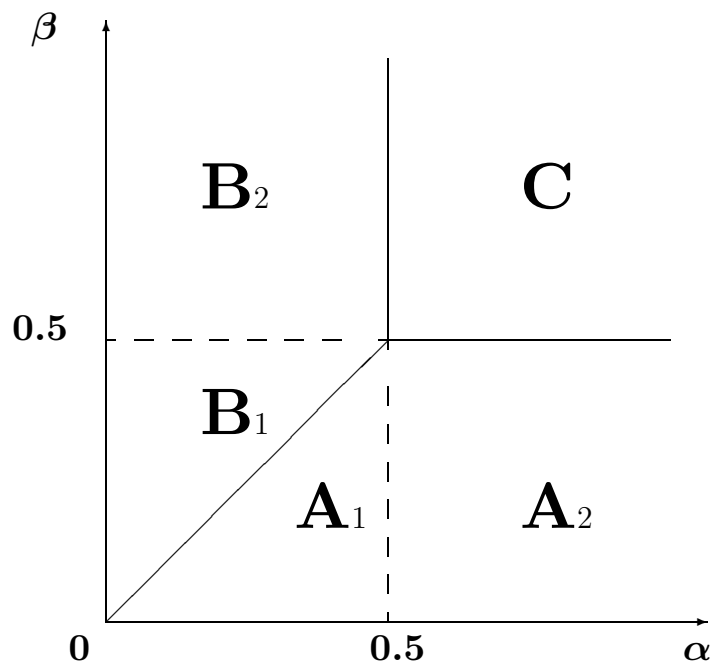
\newpage
\newpage
\quad\\[-1.5cm]
\begin{table}
\lineup
\caption[Asymmetric diffusion model: Dependence of the extrapolated gap $E_G$
on $\alpha$ and $\beta$.]
{Asymmetric diffusion model: Dependence of the extrapolated gap $E_G$
on $\alpha$ and $\beta$.
Negative values indicate that errors are of the order of $10^{-5}$.}
\footnotesize\begin{tabular}{@{}llllllllllll}
\br
$\beta\setminus \alpha$&0.1&0.2&0.3&0.4& &0.5&0.6&0.7&0.8&0.9&1.0\\\mr
1.0&0.12573 &0.04014 &0.01017 &0.00119 &\phantom{xx} &0.00000 &0.00001 &\-0.0000
2 &\-0.00001 &\-0.00000 &0.00000 \\
\\
0.9&0.12575 &0.04014 &0.01021 &0.00118 &\phantom{xx} &\-0.00003 &0.00000 &\-0.00
004 &\-0.00001 &\-0.00000 \\
\\
0.8&0.12573 &0.04013 &0.01019 &0.00118  &\phantom{xx}&\-0.00001 &0.00000 &\-0.00
004 &\-0.00004 \\
\\
0.7&0.12587 &0.04012 &0.01019 &0.00114  &\phantom{xx}&\-0.00003 &\-0.00002 &\-0.
00004 \\
\\
0.6&0.12017 &0.04016 &0.01030 &0.00120  &\phantom{xx}&\-0.00001 &\-0.00001 \\
\\
0.5&0.09999 &0.03331 &0.00861 &0.00110  &\phantom{xx}&0.00000 \\
\\
& & & & & \\
0.4&0.07213 &0.02018 &0.00330 &\-0.00005 \\
\\
0.3&0.04173 &0.00678 &\-0.00001 \\
\\
0.2&0.01428 &\-0.00001 \\
\\
0.1&\-0.00000 \\
\br
\end{tabular}
\end{table}
\quad\\[-1.5cm]
\begin{table}
\caption[Asymmetric diffusion model: Extrapolated exponents
of the first excited state.]
{Asymmetric diffusion model: Extrapolated exponents
of the first excited state.}
\lineup
\footnotesize\begin{tabular}{@{}lllllllllll}
\br
$\beta\setminus \alpha$&0.1&0.2&0.3&0.4&0.5&0.6&0.7&0.8&0.9&1.0\\\mr
1.0&  & & && 1.5000 & 1.470 & 1.456 & 1.480 & 1.4999 & 1.522 \\
\\
0.9&  & & && 1.499 & 1.452 & 1.503 & 1.4994 & 1.4999 \\
\\
0.8& &  & &  & 1.4997 & 1.430 & 1.436 & 1.499 \\
\\
0.7&  & &  &  & 1.462 & 1.405 & 1.499 \\
\\
0.6&  & &  &  & 1.434 & 1.377 \\
\\
0.5& &  &  &  & 1.49999 \\
\\
0.4&  &  &  & 2.01 \\
\\
0.3&  &  & 1.98 \\
\\
0.2&  & 2.010 \\
\\
0.1& 1.992 \\
\br
\end{tabular}
\end{table}

\begin{table}
\caption{Extrapolated and predicted values
of the energy gaps, computed from the finite size spectrum
of $-H$ for $\alpha=\beta=0.5$, $q=0$, $p=1$ and $r=0$. }
\begin{indented}
\lineup
\item[]\begin{tabular}{@{}llllll}
\br
$m$&\multicolumn{2}{c}{$R^2m^2\qquad$}&
\multicolumn{2}{c}{$2R^2my\qquad$}&
\multicolumn{1}{c}{$2R^2mx$}\\\mr

&prediction&extrapolated&prediction&extrapolated&extrapolated\\\mr
\quad&&\phantom{xxxxxxxxxxxxx}&&\phantom{xxxxxxxxxxxxx}&\\
$\frac{1}{2}$&\00.317\,648&\00.317\,63(2)&0.182\,104&
0.182\,104&\-0.120\,70(3)\\
1&\01.270\,594&\01.270\,(6)&0.364\,209&0.364\,2(1)&\-0.241\,4(1)\\
$\frac{3}{2}$&\02.858\,836&\02.858\,(9)&0.546\,313&0.546\,313&\-0.362\,10(7)\\
2&\05.082\,375&\05.082\,(3)& 0.728\,418&0.728\,(4)&\-0.482\,8(2)\\
$\frac{5}{2}$&\07.941\,211&\07.9(4)&0.910\,522&0.9(1)&\-0.60(4)\\
3&11.435\,343&11.4(8)&1.092\,627&1.0(8)&\-0.72(3)\\
\br
\end{tabular}
\end{indented}
\end{table}
\quad\\[-1.5cm]
\begin{table}
\begin{center}
\caption[Like table III for $r\geq 0$ in the sectors
$m=0$ and $m=\frac{1}{2}$.]
{Data corresponding to (4.13) from the extrapolation of the spectra
of $-H$ for $\alpha=\beta=0.5$, $q=0$, $p=1$ with $2\leq N\leq21$;
predicted degeneracies are given in
square brackets;
the predicted values of $2R^2mx$ are determined from the extrapolation
of the value $2R^2mx$ for $m=\frac{1}{2}$ and $r=0$.}
\lineup
\footnotesize\begin{tabular}{@{}llllllll}
\br
&&$m=0$&&&&$m=\frac{1}{2}$&\\\mr
$r$&$R^2m^2+r$&$2R^2mx$&$2R^2my$&
$r$&$R^2m^2+r$&$2R^2mx$&$2R^2my$\\\mr
&&&&&&&\\
1 [1]&1.000\,000   &&&0 [1]&0.317\,63(2)   &\-0.120\,70(3)   &0.182\,104   \\
&&&&&&&\\
2 [2]&1.999\,99(7)   &0.000\,0(0)   &0.000\,00(1)   &1 [1]&1.317\,65(7)   &\-0.
120\,702   &0.182\,104   \\
&&&&&&&\\
3 [3]&3.000\,0(0)   &&&2 [2]&2.317\,(4)   &\-0.120\,(7)   &0.182\,1(1)   \\
&2.999\,(8)   &\-0.000\,(1)   &\-0.000\,(0)   &&2.317\,6(8)   &\-0.120\,70(2)
 &0.182\,104   \\
&&&&&&&\\
4 [5]&4.000\,0(0)   &&&3 [3]&3.3(2)   &\-0.12(0)   &0.182\,(1)   \\
&4.00(0)   &0.000\,(1)   &\-0.000\,(0)   &&3.317\,(7)   &\-0.120\,7(0)   &0.182
\,1(0)   \\
&4.0(0)   &0.00(3)   &\-0.00(0)   &&3.31(7)   &\-0.120\,6(8)   
&0.182\,1(1)   \\ &&&&&&&\\
5 [7]&4.99(6)   &&&4 [5]&4.3(2)   &\-0.1(3)   &0.18(4)   \\
&5.00(0)   &0.00(0)   &0.000\,(0)   &&4.31(8)   
&\-0.120\,(7)   &0.182\,(1)   \\
&5.0(0)   &0.0(0)   &\-0.00(1)   &&4.317\,(7)   &\-0.120\,(8)   &0.182\,0(8)
\\
&&&&&4.31(6)   &\-0.12(2)   &0.18(2)   \\
6&5.99(7)   &&&&4.31(7)   &\-0.11(8)   &0.182\,(0)   \\
&6.0(1)   &0.0(0)   &\-0.00(1)   &&&&\\
&6.0(0)   &0.00(1)   &\-0.00(0)   &5 [7]&5.31(7)   &\-0.121\,(4)   &0.182\,0(8)
   \\
&6.0(0)   &0.0(0)   &\-0.00(1)   &&5.317\,4(8)   &\-0.120\,(7)   &0.182\,(1)
\\
&&&&&5.31(4)   &\-0.12(2)   &0.18(2)   \\
7&7.0(1)   &&&&5.31(7)   &\-0.12(1)   &0.18(2)   \\
&6.99(9)   &\-0.0(0)   &\-0.000\,(0)   &&5.3(2)   &\-0.1(3)   &0.18(4)   \\
&7.0(1)   &0.0(0)   &\-0.00(1)   &&5.3(2)   &\-0.11(5)   &0.1(8)   \\
&&&&&&&\\
8&8.0(0)   &&&6&6.3(1)   &\-0.11(7)   &0.182\,(1)   \\
&8.0(2)   &&&&6.3(4)   &\-0.1(1)   &0.18(1)   \\
&8.0(1)   &0.0(0)   &\-0.00(1)   &&&&\\
&8.(1)   &0.(0)   &\-0.00(5)   &&&&\\
&8.0(1)   &0.0(0)   &\-0.00(1)   &&&&\\
\mr
prediction
&0.000\,000$+r$&0.000\,000&0.000\,000&&0.317\,648$+r$&\llap{$[-$}0.120 70 ]&0.1
82\,104\\\br
\end{tabular}
\end{center}
\end{table}
\begin{table}
\caption{Dependence of the free surface energy $f_{\infty}$ on $\alpha,
\beta$.}
\begin{center}
\lineup
\footnotesize\begin{tabular}{@{}llllllll}
\br
$\beta\setminus \alpha$&1.5&1.3&1.1&0.9&0.7&0.5&0.3\\\mr

\\ 0.1&0.4959(9)&0.30(9)&0.123(1)&\-0.06(4)&\-0.25(6)&\-0.4631(0)&\-0.71(0)
\\ 0.3&0.9086(1)&0.722(0)&0.535(8)&0.348(6)&0.1564(6)&\-0.050216(5)
\\ 0.5&1.15624(0)&0.96957(1)&0.78344(0)&0.596234(5)&0.4040372
\\ 0.7&1.362(8)&1.176(2)&0.9900(9)&0.8028(9)
\\ 0.9&1.555(0)&1.368(4)&1.18(2)
\\ 1.1&1.742(2)&1.555(6)
\\ 1.3&1.928(3)\\\br
\end{tabular}
\end{center}
\end{table}
\begin{table}
\caption[Central charges for $\alpha=\beta$.]{Extrapolated central charges.
The value of $R^2x$ is determined from
$E_0^1$ .}
\begin{indented}
\lineup
\item[]\begin{tabular}{@{}lllll}
\br
$\alpha=\beta$&$f_{\infty}$&$R^2x$&$c$&$c'$\\\mr
\\0.1&\-1.12(3)&0.793(8)&12.9(8)&1.0(8)
\\0.2&\-0.6(2)&0.506(0)&5.8(3)&1.0(0)
\\0.3&\-0.297(8)&0.3339(8)&3.1(1)&1.0(1)
\\0.4&\-0.0336(9)&0.2126(4)&1.84(8)&0.99(4)
\\0.5&0.19737&0.1207(0)&1.2752(8)&1.0000(8)
\\0.6&0.40965&0.0486(5)&1.044(4)&0.999(7)
\\0.7&0.610(7)&\-0.008(7)&1.00(0)&0.99(9)
\\0.8&0.8049(7)&\-0.054(6)&1.04(8)&0.99(2)
\\0.9&1.1829(2)&\-0.0913(7)&1.2(7)&1.1(2)
\\1.0&1.369(5)&\-0.1(2)&1.3(8)&1.1(1)\\\br
\end{tabular}
\end{indented}
\end{table}
\begin{table}
\caption{Dependence of $R^2x$ on $\alpha$ and $\beta$.}
\begin{indented}
\lineup
\item[]\begin{tabular}{@{}llllllll}
\br
$\beta\setminus \alpha$&1.5&1.3&1.1&0.9&0.7&0.5&0.3\\\mr

\\ 0.1&\-0.29(9)&\-0.30(9)&\-0.32(5)&\-0.35(2)&\-0.39(3)&\-0.45900&\-0.565(6)
\\ 0.3&\-0.06(8)&\-0.07(8)&\-0.09(4)&\-0.12(1)&\-0.1(6)&\-0.22735
\\ 0.5&0.0379(7)&0.02806&0.01165&\-0.01467&\-0.05601
\\ 0.7&0.102(5)&0.092(6)&0.076(2)&0.049(9)
\\ 0.9&0.14(3)&0.13(3)&0.11(7)
\\ 1.1&0.17(0)&0.16(0)
\\ 1.3&0.18(6)\\\br
\end{tabular}
\end{indented}
\end{table}
\end{document}